\documentclass[aps,floats,twocolumn,prl,showpacs,nofootinbib,reprint]{revtex4-2}

\usepackage{amssymb}
\usepackage{amsmath}

\usepackage{graphicx}
\usepackage{graphics}
\usepackage{dcolumn}
\usepackage{color}
\usepackage{rotate}
\usepackage{fancyhdr}
\usepackage{subfigure}
\usepackage{ulem}

\usepackage{xcolor}
\usepackage[hidelinks]{hyperref}

\definecolor{TurkishBlue}{HTML}{144893}
\hypersetup{colorlinks=true,citecolor=TurkishBlue,linkcolor=TurkishBlue,urlcolor=TurkishBlue}

\def\be{\begin{equation}}
\def\ee{\end{equation}}
\def\bea{\begin{eqnarray}}
\def\eea{\end{eqnarray}}
\def\ba{\begin{eqnarray}}
\def\ea{\end{eqnarray}}

\def\VEV#1{\left\langle #1 \right\rangle}

\definecolor{darkred}{rgb}{.743,0,0}

\begin{document}

\title{Phenomenology of a vector-field-induced (and possibly parity breaking) compensated isocurvature perturbation}

\author{Eleonora Vanzan$^{1,2}$, Marc Kamionkowski$^3$, and Selim~C.~Hotinli$^{3,4}$}
\affiliation{$^1$Dipartimento di Fisica Galileo Galilei,
Universit\'a di Padova, I-35131 Padova, Italy}
\affiliation{$^2$INFN Sezione di Padova, I-35131 Padova, Italy}
\affiliation{$^3$William H.\ Miller III Department of Physics and Astronomy, Johns Hopkins University, 3400 N.\ Charles St., Baltimore, MD 21218}
\affiliation{$^4$Perimeter Institute for Theoretical Physics, 31 Caroline St N, Waterloo, ON N2L 2Y5, Canada}

\date{\today}

\begin{abstract}
It is natural to wonder whether there may be observational relics of new fundamental fields, beyond the inflaton, in large scale structure.  Here we discuss the phenomenology of a model in which compensated isocurvature perturbations (CIPs) arise through the action of a primordial vector field that displaces dark matter relative to baryons.  The model can be tested best by kinematic-Sunyaev-Zeldovich tomography, which involves the cross-correlation of cosmic microwave background and galaxy surveys, with next-generation observatories.  There are also signatures of the vectorial nature of the new field that may be detectable in forthcoming galaxy surveys, but the galaxy survey cannot alone indicate the presence of a CIP.  Models that induce a parity breaking four-point correlation in the galaxy distribution are also possible.
\end{abstract}

\maketitle

%\tableofcontents

Cosmologists have long wondered whether there may be fossils of new fundamental fields, beyond the inflaton, in large-scale structure. The majority of such work has involved a new scalar field (e.g.,~the curvaton~\cite{Lyth:2001nq}), but it is conceivable that higher-spin fields may also have been active in the early Universe.
One obvious example is the graviton field, which can be excited during standard single-field slow-roll inflation~\cite{Starobinsky:1979ty,Rubakov:1982df,Fabbri:1983us,Abbott:1984fp} and the higher-order correlations that it may induce in the cosmological mass distribution through cross-correlation with the inflaton~\cite{Maldacena:2002vr}. 
Other possibilities include higher-order correlations in the matter distribution, induced by coupling of primordial perturbations to a primordial magnetic field during inflation~\cite{Caldwell:2011ra}. A general description of the effects of couplings of an inflaton to spin-1 or spin-2 fields was provided in Ref.~\cite{Jeong:2012df}, and related ideas have recently been developed~\cite{Cabass:2022oap,Cabass:2022rhr,Orlando:2022rih,Niu:2022fki,Creque-Sarbinowski:2023wmb,Tong:2022cdz,Vanzan:2024tiq}, with the addition of parity-breaking physics, to account for recently reported evidence for parity violation in the four-point correlation function in galaxy surveys~\cite{Philcox:2022hkh,Hou:2022wfj}.

In this paper we describe a toy model in which the effects of a primordial vector field are imprinted into large-scale-structure observables in a novel way: compensated isocurvature perturbations (CIPs)~\cite{Holder:2009gd,Gordon:2009wx,Grin:2011tf,Grin:2011nk}---perturbations to the baryon and dark-matter densities of equal but opposite amplitude---are induced by displacements of baryons and dark matter that are described by a transverse vector field. If we start with a perfectly homogeneous Universe and then induce a transverse displacement of the baryons relative to the dark matter, the Universe remains homogeneous. The CIP in this model is thus induced only if the Universe has some nonzero adiabatic density perturbation (as it does). As recent work has shown~\cite{Kumar:2022bly}, the sensitivity to CIPs will improve by several orders of magnitude with forthcoming kSZ-tomography surveys, as compared to current CMB and galaxy-survey bounds.
It is thus reasonable to consider new observables, such as those we describe below, that can be used to characterize such CIPs.

The CIP induces perturbations to the baryon (b) and cold-dark-matter (c) densities,
\begin{eqnarray}
     \rho_b({\bf x}) &=& \bar \rho_b[1 +  \Delta({\bf x})], \nonumber \\
     \rho_c({\bf x}) &=& \bar \rho_c[1 - f_b \Delta({\bf x})],
\label{eqn:cipperturbation}
\end{eqnarray}
in terms of a CIP field $\Delta({\bf x})$, where $\bar \rho_i$ are the mean densities, and $f_b=\bar \rho_b/\bar \rho_c$.

Here we surmise that the CIP is induced by a displacement ${\bf A}({\bf x})$ of the baryons and dark matter, in a way that conserves the total-matter density.  More precisely, if $\rho_m({\bf x}) =\rho_b({\bf x}) + \rho_c({\bf x})$ is the total nonrelativistic-matter density, then the new matter densities in the presence of the displacement field are
\begin{eqnarray}
     \left. \rho_b({\bf x})\right|_{\bf A} &=& \frac{f_b}{1+f_b} \rho_m({\bf x}+{\bf A}) \simeq \rho_b({\bf x}) + {\bf A} \cdot  \nabla \rho_b({{\bf x}}), \nonumber \\ 
     \left. \rho_c({\bf x})\right|_{\bf A} &=& \frac{1}{1+f_b}\rho_m({\bf x} - f_b{\bf A}) \simeq \rho_c({\bf x}) -  {\bf A} \cdot \nabla \rho_b({{\bf x}}).\nonumber \\
\end{eqnarray}
Here we have assumed that there is a curvature perturbation in place with fractional nonrelativistic density perturbation $\delta_m({\bf x})=\delta\rho_m({\bf x})/\bar \rho_m = \delta_b({\bf x}) =\delta_c({\bf x})$. The prediction of the theory is thus that
\begin{equation}
      \Delta({{\bf x}}) = {\bf A}({\bf x}) \cdot \nabla\delta_m({{\bf x}}).
\label{eqn:defining}
\end{equation}
Most generally, the displacement field ${\bf A}({\bf x})$ can be decomposed into a longitudinal and transverse part. The longitudinal component, however, can be written as the gradient of a scalar function and thus reproduce the standard CIP phenomenology. We therefore consider only a transverse (i.e.,~$\nabla \cdot {\bf A}({\bf x})=0$) vector field in what follows.

Here we describe how to characterize a CIP field from a transverse field ${\bf A}({\bf x})$ in the event that $\delta_m({\bf x})$ and $\Delta(\bf x)$ are measured. We do not attempt here to embed this phenomenological model in a more complete theory. We surmise, though, that this might be accomplished in variants of models of asymmetric dark matter~\cite{Kaplan:2009ag,Zurek:2013wia} in which baryons and dark matter are in some representation that is charged under a gauge group related to ${\bf A}({\bf x})$.

The vector field is taken to be a random field with power spectrum $P_A(k)$ defined so that the cartesian components of the Fourier transform ${{\bf A}({\bf k})}$ satisfy~\cite{Adi:2023doe} (assuming for now parity conservation---more on parity breaking later),
\begin{equation}
     \VEV{A_i^*({\bf k}) A_j({\bf k}')} = (2\pi)^3 \delta_D({\bf k}-{\bf k}') \left( \delta_{ij} - k_i k_j/k^2 \right) P_A(k).
\label{eqn:Acorrelations}     
\end{equation}     

\begin{figure}[htbp]
    \centering
    \includegraphics[width=.5\textwidth]{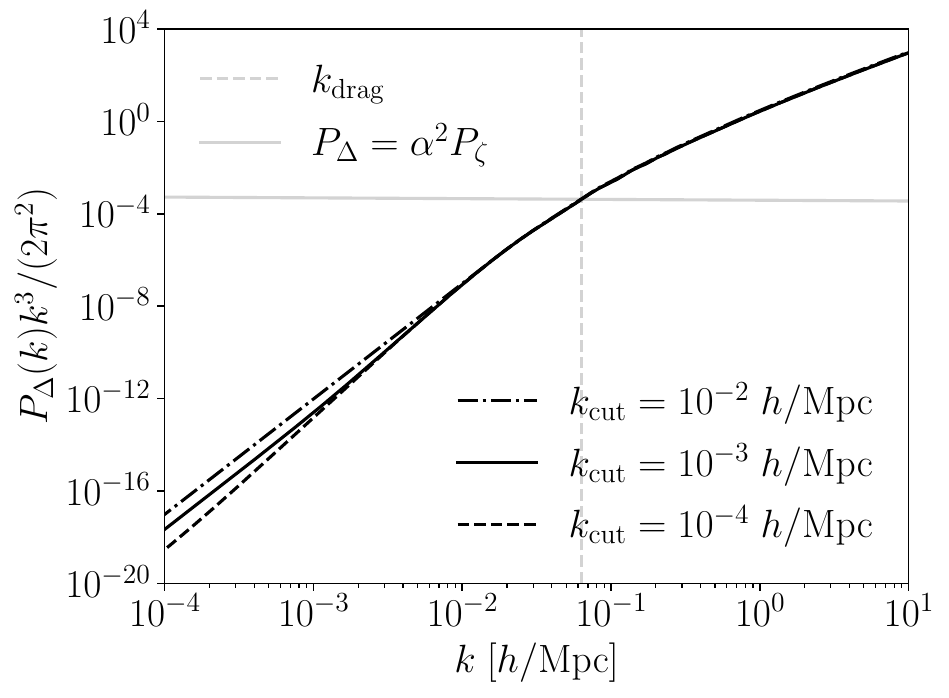}
    \caption{The CIP power spectrum with a power-law $P_A(k)=\mathcal{A}k^n \theta_H(k-k_{\rm cut})$ power spectrum for the vector field, for different cutoff scales; the results for $n=-6$, $n=-5$, and $n=-7$ are similar. The normalization $\mathcal{A}$ is obtained by matching the amplitude of the CIP power spectrum at $k=0.05$ $h$~Mpc$^{-1}$ to the upper bound, obtained from the CMB for a scale-invariant CIP spectrum.}
    \label{fig:fig1}
\end{figure}

Given Eq.~(\ref{eqn:defining}), the CIP field will be a convolution in Fourier space. Assuming that the vector field is uncorrelated with the total matter density fluctuations, this yields a CIP power spectrum $P_\Delta(k)$, defined by $\VEV{\Delta^*({\bf k}) \Delta({\bf k}')} = (2\pi)^3 \delta_D({\bf k}-{\bf k}') P_\Delta(k)$,  given by
\begin{equation}
     P_\Delta(k) = k^2 \int \frac{d^3{\bf k}_1}{(2\pi)^3} (1-\mu^2) P_m\left(\left| {\bf k}-{\bf k}_1 \right| \right) P_A(k_1),
\label{eqn:convolution}
\end{equation}
with $\mu=\hat{k}\cdot\hat{k}_1$ and $P_m(k)$ is the matter power spectrum.  The result of the convolution, with the standard $\Lambda$CDM matter power spectrum, is plotted in Fig.~\ref{fig:fig1}, for a power-law power spectrum $P_A(k)=\mathcal{A}k^n \theta_H(k-k_{\rm cut})$ cut off at a wavenumber $k_{\rm max}$.

In order to assess the prospects for this model to be probed by future measurements, we fix the normalization $\mathcal{A}$ of the vector-field power spectrum so that the induced CIP perturbation takes on the highest amplitude consistent with current constraints.  Strictly speaking, CMB constraints to CIPs have been obtained for a CIP power spectrum $P_\Delta(k) \propto k^{-3}$, which is different from that shown in Fig.~\ref{fig:fig1}.  However, the CMB bound is weighted primarily by Fourier modes near $k\sim0.05$~Mpc$^{-1}$.  We therefore take 
\begin{equation}
    P_{\Delta}(k) \lesssim \frac{2\pi^2 \alpha^2 A_s}{k^3},
\end{equation}
at $k=0.05$~Mpc$^{-1}$ with $\alpha \sim 450$ \cite{Barreira:2020lva}, but the actual upper bound to ${\cal A}$ may be a bit higher or lower.

Suppose now that a nonzero CIP perturbation has been detected, and that we would like to figure out if it, or any part of it, arises from a vector field through Eq.~(\ref{eqn:defining}).  To do so, we derive a quadratic estimator for the components of ${{\bf A}({\bf x})}$ as follows.  Given a realization of the ${\bf A}$ field, the Fourier components of the CIP and matter perturbation have expectation values,
\begin{eqnarray}
     \VEV{ \delta_m({\bf k}) \Delta({\bf k}') } &=& \sum_{{\bf k}_1} {\bf A}({\bf k}' - {\bf k}_1) \cdot i{\bf k}_1 \VEV
     {\delta_m({\bf k})\delta_m({\bf k}_1)} \nonumber \\
     &=& -i{\bf k} \cdot {\bf A}({\bf k} + {\bf k}') P_m(k).
\end{eqnarray}
This implies that every pair $\delta_m({\bf k})$ and $\Delta({\bf k}')$ with ${\bf k}+ {\bf k}' = {\bf K}$ provides an estimator,
\begin{equation}
     {\bf k} \cdot \widehat{{\bf A}({\bf K})}_{{\bf k},{\bf k}'} =
     i \frac{\delta_m({\bf k}) \Delta({\bf k}')}{P_m(k)}.
\label{eqn:kkprimeestimator}
\end{equation}
Strictly speaking, Eq.~(\ref{eqn:kkprimeestimator}) provides an estimator only for the projection of ${\bf A}({\bf K})$ along the component of ${\bf k}$ parallel to ${\bf A}$.  To be a bit more precise, suppose we align the $\hat z$ direction with the direction of the wavevector ${\bf K}$ for the Fourier mode under consideration.  We can then choose two other directions $\hat x$ and $\hat y$ orthogonal to $\hat z$ and take $\theta$ to be the angle between ${\bf K}$ and ${\bf k}$ and $\phi$ to be the azimuthal angle in the $x$-$y$ plane.  We then arrive at a vector-valued estimator,
\begin{equation}
     \widehat{{\bf A}({\bf K})}_{{\bf k},{\bf k}'} =
     i \frac{\delta_m({\bf k}) \Delta({\bf k}')}{k \sin\theta P_m(k)}
     (\cos^{-1}\phi,\sin^{-1}\phi),
\label{eqn:kkprimeestimatorvector}
\end{equation}     
for the two nonzero components of ${\bf A}({\bf K})$.  We now take as our null hypothesis a CIP power spectrum $P_\Delta(k)$, but ${\bf A}({\bf x})=0$.  The variance of Eq.~(\ref{eqn:kkprimeestimatorvector}) under the null hypothesis is
\begin{equation}
     \VEV{ \left|\widehat{{\bf A}({\bf K})}_{{\bf k},{\bf k}'}
     \right|^2 } = \frac{ P_m^{\rm tot}(k) P_\Delta^{\rm tot}(k')}{k^2 \sin^2\theta P_m(k)^2},
     %\VEV{ \left|\widehat{{\bf A}({\bf k})}_{{\bf k},{\bf k}'}
     %\right|^2 } = \frac{P_\Delta(k')}{k^2 \sin^2\theta P_m(k)}.
\end{equation}
$P^{\rm tot}$ are the observed power spectra including the noise.
We can then add the estimators built from all $\delta_m({\bf k})$-$\Delta ({\bf k}')$ pairs with ${\bf k}+ {\bf k}' = {\bf K}$ with inverse-variance weighting to obtain the minimum-variance estimator,
\begin{eqnarray}
     \widehat{{\bf A}({\bf K})} = && P_n^A(K) i
     \sum_{{\bf k}+{\bf k}'={\bf K}} \frac{k \sin\theta P_m(k)}{ P_m^{\rm tot}(k) P_{\Delta}^{\rm tot}(k')} \nonumber \\
     &&  \times \delta_m({\bf k})\Delta({\bf k}') (\cos^{-1}\phi,\sin^{-1}\phi).\nonumber \\
\label{eqn:optimalestimator}
\end{eqnarray}
where the inverse variance is, given our shorthand $\sum_{{\bf k}}$ for $  \int d^3{\bf k}/(2\pi)^3$, the noise power spectrum,
\begin{eqnarray}
     \left[P_n^A(K) \right]^{-1}\!\! &=& \frac{1}{4\pi^2} \int_{k_{\rm min}}^{k_{\rm max}} \, dk\, k^4\, \frac{P_m(k)^2}{P_m^{\rm tot}(k)} \nonumber \\
     && \times \int_{-1}^1
     d\mu \,  \frac{1-\mu^2}{P_\Delta^{\rm tot}\left( \sqrt{K^2+k^2 -
     2 K k\mu}\right)},\nonumber \\
\label{eqn:noisepowerspectrum}  
\end{eqnarray}
with $k_{\rm min}=2\pi/V^{1/3}$ in terms of the volume $V$ of the survey.

If the amplitudes $\widehat{{\bf A}({\bf K})}$ arise as a realization of a random field with power spectrum $P(K)=\mathcal{A}P_f(K)$, for some fiducial power spectrum $P_f(K)=K^{n_f}$ and amplitude $\mathcal{A}$, then each Fourier mode ${\bf K}$ provides an estimator,
\begin{equation}
    \widehat{\mathcal{A}}_K = \frac{1}{P_f(K)} \left[  \left| \widehat{{\bf A}({\bf K})} \right|^2 -P_n^A(K) \right],
\end{equation}
for that amplitude with variance $2\left[P_n^A(K)\right]^2 / \left[P_f(K)\right]^2$.  The minimum-variance estimator for $\mathcal{A}$ is then,
\begin{equation}
    \widehat{\mathcal{A}} = \sigma_{\mathcal{A}}^2 \sum_{{\bf K}} \frac{P_f(K)}{2P_n(K)^2} \left[  \left| \widehat{{\bf A}({\bf K})} \right|^2 -P_n^A(K) \right],
\end{equation}
which has a variance,
\begin{equation}
    \sigma_{\mathcal{A}}^{-2} = \sum_{{\bf K}} \frac{\left[P_f(K)\right]^2}{2 \left[P_n^A(K)\right]^2},
\end{equation}
where again $K$ ranges from $k_{\rm min}$ to $k_{\rm max}$.

In order to estimate the variance $\sigma_{\mathcal{A}}^2$ and thus the smallest detectable $\mathcal{A}$, we need the expected noise contributions to $P_m(k)$ and $P_\Delta(k)$ from kSZ tomography.
\newcommand{\muLOS}{\mu_{\rm los}}
kSZ tomography reconstructs the radial component of the peculiar-velocity field from which the total-matter density field is reconstructed.  The noise power spectrum expected for DESI and CMB-S4 is shown in Fig.~5 in Ref.~\cite{Smith:2018bpn} and can be approximated by,
\begin{equation}
    P_{mm}^n(k,\muLOS) = \frac{1}{\muLOS^2} a \left( \frac{k}{k_0} \right)^b ,
\label{eqn:kSZnoise}
\end{equation}
with $k_0=0.05 \ {\rm Mpc}^{-1}$, and $a=36091 \ ({\rm Mpc}/h)^3$, $b=1.99$, and where $\muLOS$ is the cosine of the angle that the mode makes with the line of sight.

For the CIP field, we suppose that the fractional galaxy-density perturbation can be modeled as~\cite{Barreira:2020lva},
\begin{equation}
    \delta_g({\bf x}) = b_m \delta_m({\bf x}) +b_{\Delta} \Delta({\bf x}),
\label{eqn:deltag}    
\end{equation}
with bias parameters $b_m$ and $b_\Delta$.  We then  obtain \cite{Kumar:2022bly} $\widehat{\Delta} = \left( \delta_g-b_m\delta_m \right)/b_{\Delta}$, which then has a noise power spectrum,
\begin{equation}
    P_{\Delta\Delta}^n({\bf k}) = \frac{1}{b_{\Delta}^2} \left[ \frac{1}{\bar{n}_g} +b_m^2 P_{mm}^n({\bf k}) \right],
\end{equation}
where $\bar{n}_g$ is the galaxy number density.

Strictly speaking, Eq.~(\ref{eqn:noisepowerspectrum}) is written assuming that the matter and CIP reconstruction noises are isotropic, whereas they are in fact dependent on $\muLOS$, as seen above.  However, a rough estimate can be obtained by assuming that all three components (rather than just the line-of-sight component) of the peculiar-velocity field can be reconstructed and then by increasing the noise by a factor of 3.

The dependence in Eq.~(\ref{eqn:deltag}) on the CIP field and the dependence in Eq.~(\ref{eqn:defining}) of the CIP field on the vector field imply a characteristic non-gaussianity in the galaxy-density field that can also be sought.  There is a correlation between two modes $\delta_g({\bf k})$ and $\delta_g({\bf k'})$, with ${\bf k} \neq {\bf k}'$,
\begin{equation}
    \VEV{\delta_g({\bf k}) \delta_g(\bf k')} = -i {\bf A}({\bf k}+{\bf k}') \cdot \left[ {\bf k} P_m(k) + {\bf k}' P_m(k') \right].
\label{eqn:vectorfossil}    
\end{equation}
This then implies that every such pair with ${\bf k}+{\bf k}' = {\bf K}$ provides an estimator,
\begin{eqnarray}
     \widehat{{\bf A}({\bf K})}_{{\bf k},{\bf k}'} = &&
     i \frac{\delta_g({\bf k}) \delta_g({\bf k}')}{b_{\Delta} b_m k \sin\theta \left[ P_m(k)-P_m(k') \right]} \nonumber \\
     &&  \times \left( \cos^{-1}\phi, \sin^{-1}\phi \right) ,
\end{eqnarray}
with variance
\begin{equation}
    \VEV{ \left|\widehat{{\bf A}({\bf K})}_{{\bf k},{\bf k}'} \right|^2 } =  \frac{P_{gg}^{\rm tot}(k) P_{gg}^{\rm tot}(k')}{  b_m^2 b_\Delta^2 k^2\sin^2\theta \left[ P_m(k)-P_m(k') \right]^2 }.
\end{equation}
Following the same reasoning that led to Eq.~(\ref{eqn:noisepowerspectrum}), we find that the noise for ${\bf A}({\bf K})$ obtained with a galaxy survey will be
\begin{eqnarray}
     \left[P_n^A(K) \right]^{-1}  &=& \frac{b_m^2 b_\Delta^2}{8\pi^2} \int_{k_{\rm min}}^{k_{\rm max}} \, dk\, k^4\, \int_{-1}^1\, d\mu\, (1-\mu^2) \nonumber \\
     && \times \frac{\left[P_m(k) - P_m\left(\sqrt{K^2+k^2 -
     2 K k\mu}  \right) \right]^2}{P_{gg}^{\rm tot}(k) P_{gg}^{\rm tot}\left( \sqrt{K^2+k^2 -
     2 K k\mu}\right)}.\nonumber \\
\label{eqn:galaxynoisepowerspectrum}    
\end{eqnarray}
The additional factor of two in the denominator is there to account for double counting of equivalent ${\bf k}$-${\bf k}'$ pairs.

\begin{figure*}[htbp]
    \centering
    \includegraphics[width=\textwidth]{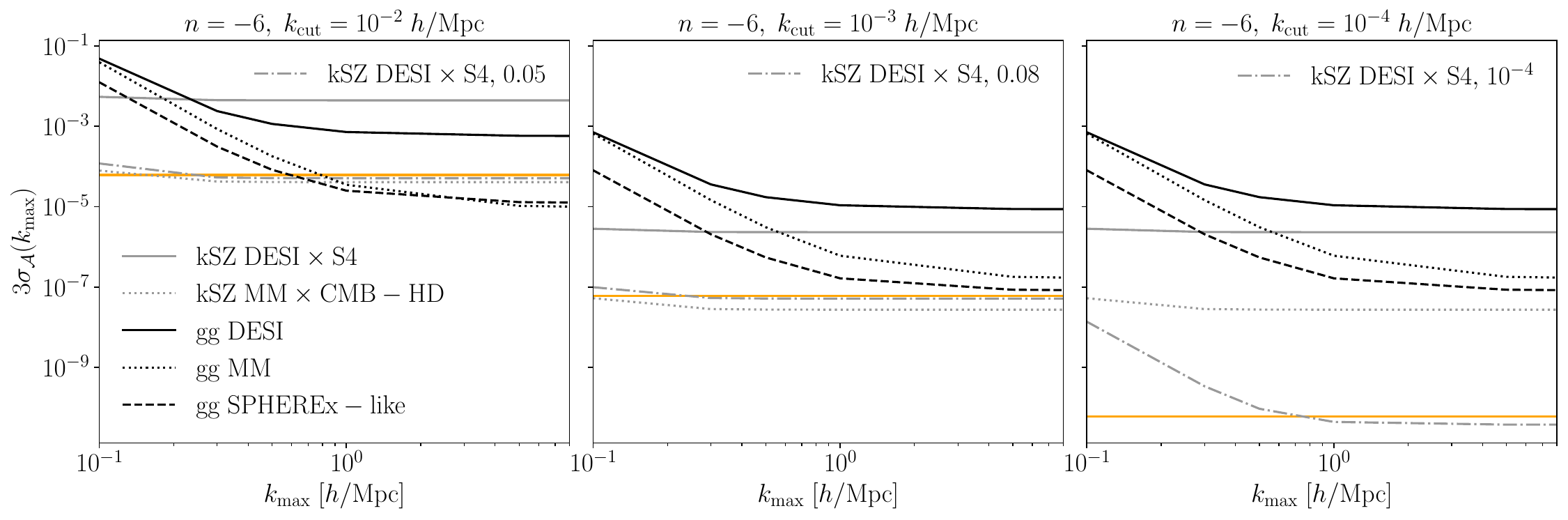}
    \caption{Detection threshold $3\sigma_{\mathcal{A}}$ for a vector-field power spectrum $P_A(k)=\mathcal{A}k^n \theta_H(k-k_{\rm cut})$, at fixed $n=-6$, for three different choices of the wavenumber cutoff. This is to be compared to the value of the maximum amplitude $\mathcal{A}$ (orange lines) consistent with current CIP constraints. DESI specifications are taken from Ref.~\cite{Smith:2018bpn}: the mean redshift is $z=0.75$, the survey volume $V=36 \ ({\rm Gpc}/h)^3$, galaxy number density $n_g = 5.5 \times 10^{-4} \ (h/{\rm Mpc})^3$, and galaxy bias $b_m=1.51$. The reconstruction noise from kSZ tomography for DESI$\times$CMB-S4 is that given in Eq.~(\ref{eqn:kSZnoise}), setting $\muLOS=1$ and multiplying the final result by $\sqrt{3}$ as a rough estimate of anisotropic reconstruction effects. In order to investigate the prospects for improvements beyond DESI$\times$CMB-S4, the kSZ reconstruction noise amplitude $a$ is multiplied by an improvement factor 0.05, 0.08, $10^{-4}$ from left to right (gray, dash-dot lines), or decreased to $a=1665 \ ({\rm Mpc}/h)^3$ that we estimate for MegaMapper (MM)$\times$CMB-HD (gray, dotted lines).  We then see that for the two leftmost panels the expected signal may be within reach of MegaMapper$\times$CMB-HD, with $z=0.75$, $V=31 \ ({\rm Gpc}/h)^3$, $n_g = 0.01 \ (h/{\rm Mpc})^3$, and $b_m=1.6$.  For the galaxy-galaxy correlation, a SPHEREx-like survey (black, dashed lines) with $z=1$, $V=100 \ ({\rm Gpc}/h)^3$, $n_g = 5 \times 10^{-3} \ (h/{\rm Mpc})^3$, and $b_m=1.5$ may be able to detect these vector-field CIPs with $k_{\rm cut}\gtrsim 10^{-3} \ h/{\rm Mpc}$.}
    \label{fig:fig2}
\end{figure*}

The detection threshold $3\sigma_{\mathcal{A}}$ is plotted in Fig.~\ref{fig:fig2} for $n=-6$ and for three different choices of the cutoff; the results for $n=-5$, $n=-7$ are similar. The DESI$\times$S4 configuration is not enough to achieve detection for the cases considered; however, depending on the cutoff scale of the vector-field power spectrum, a factor-of-ten improvement in the kSZ reconstruction noise may allow detection.
For the galaxy-galaxy estimator, DESI is not sufficient, but SPHEREx~\cite{SPHEREx:2014bgr} or MegaMapper~\cite{Schlegel:2019eqc} might perform nearly as well as kSZ tomography for $k_{\rm cut}=10^{-3} \ h/{\rm Mpc}$ and up to an order of magnitude better for $k_{\rm cut}=10^{-2} \ h/{\rm Mpc}$.  Here we took the maximum CIP amplitude ${\cal A}$ to be consistent with CMB bounds as an estimate, but the actual value might turn out to be a bit higher or lower.

Given the timescales of the projects considered (SPHEREx is coming online soon), the prospects to soon seek the signatures of this model are better for galaxy surveys.  However, these vector-induced correlations in a galaxy survey, if detected, would not signal detection of CIPs.  This galaxy correlation does not indicate unambiguously a spatially varying baryon-to-dark-matter ratio, as it probes only the galaxy distribution, not the dark-matter distribution.  That would require a signal to be seen in kSZ tomography \cite{Kumar:2022bly}.

If the correlation in Eq.~(\ref{eqn:vectorfossil}) exists, then another with some other pair of modes with wavenumbers ${\bf k}_1$, ${\bf k}'_1$ with ${\bf k}_1+{\bf k}_1'=-{\bf K}$ will also exist.  This thus indicates a nonzero four-point correlation (in Fourier space) for the four modes with ${\bf k}+{\bf k}'+{\bf k}_1+{\bf k}_1'=0$ and is thus equivalent to a non-zero four-point correlation function.  In Eq.~(\ref{eqn:Acorrelations}), it was assumed that the vector-field correlations were parity conserving.  If, however, some mechanism gives rise to a preference of right-circularly polarized vector modes over left-circularly polarized modes---a chiral vector-field background---then this four-point function will be parity breaking \cite{Jeong:2012df}. The model considered here might, with a bit more development, provide an explanation for the types of parity-breaking four-point correlations discussed in Refs.~\cite{Hou:2022wfj,Philcox:2022hkh}. Such a model could be distinguished from other parity-violating models with better characterization of the four-point function.

To close, we have discussed a novel model of CIPs that are induced by a transverse-vector-valued baryon-dark-matter displacement field.  The CIP and vectorial features of the model can be probed with kSZ tomography.  The vectorial nature can be probed with the galaxy four-point correlation function.  The model provides an implementation of the vector fossil fields considered in Ref.~\cite{Jeong:2012df}, and it also allows for a new way to generate parity-breaking correlations in the galaxy distribution.  It should be interesting, in future work, to take the next steps in building and considering these models.

\smallskip
We thank Alvise Raccanelli for useful comments. This work was supported at JHU by NSF Grant No.\ 2112699, the Simons Foundation, and the Templeton Foundation. SCH was supported by the P.~J.~E.~Peebles Fellowship at Perimeter Institute for Theoretical Physics and the Horizon Fellowship from Johns Hopkins University. Research at Perimeter Institute is supported by the Government of Canada through the Department of Innovation, Science and Economic Development Canada and by the Province of Ontario through the Ministry of Research, Innovation and Science.

\bibliography{vectorcips}

\end{document}